\documentclass[journal=jpccck,manuscript=article]{achemso}
\usepackage[T1]{fontenc} 
\usepackage{amsmath}
\usepackage{amssymb}
\usepackage{natbib}
\usepackage[svgnames]{xcolor}
\usepackage[colorlinks=false,citecolor=blue,linkcolor=blue,citebordercolor=white,linkbordercolor=white]{hyperref}

\def\eq#1{{eq \ref{#1}}}

\def\Eq#1{{Equation \ref{#1}}}

\def\onlinecite#1{{ref \citenum{#1}}}

\def\fig#1{{Figure \ref{#1}}}

\def\vec#1{{\bf{#1}}}

\author{S.V. Novikov}
\email{novikov@elchem.ac.ru}
\phone{+7 495 952 24 28}
\fax{+7 495 952 53 08}
\affiliation[Frumkin Institute]
{A.N. Frumkin Institute of
Physical Chemistry and Electrochemistry, Leninsky prosp. 31,
119071 Moscow, Russia}
\alsoaffiliation[HSE]
{National Research University Higher School of Economics, Myasnitskaya Ulitsa 20, Moscow 101000, Russia}

\title[Recombination]{Bimolecular Recombination of Charge Carriers in Polar Amorphous Organic Semiconductors: Effect of Spatial Correlation of the Random Energy Landscape}

\SectionNumbersOn

\begin{document}

\begin{tocentry}
\includegraphics[width=2in]{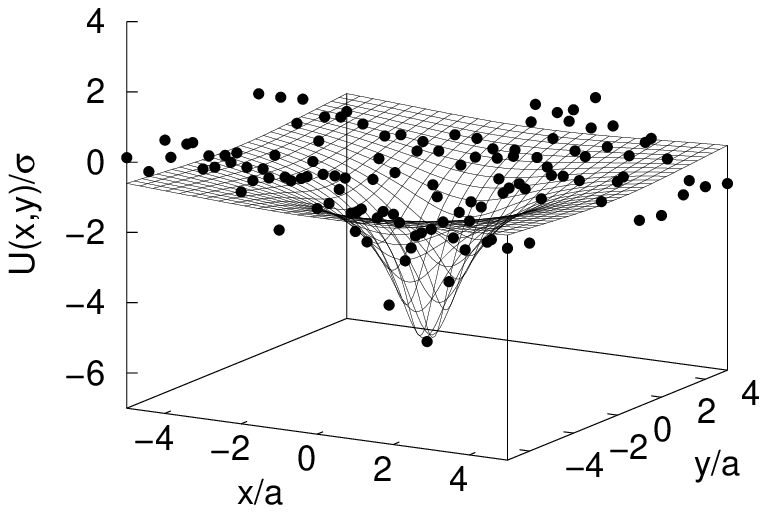}

\end{tocentry}

\begin{abstract}
We present a simple model of the bimolecular charge carrier recombination in polar amorphous  organic semiconductors where the dominant part of the energetic disorder is provided by permanent dipoles and show that the recombination rate constant could be much smaller than the corresponding Langevin rate constant. The reason for the strong decrease of the rate constant is the long range spatial correlation of the random energy landscape in amorphous dipolar materials, without spatial correlation even strong disorder does not modify the Langevin rate constant. Our study shows that the significant suppression of the bimolecular recombination could take place in homogeneous amorphous organic semiconductors and does not need large scale inhomogeneity of the material.
\end{abstract}


\newpage

\section{Introduction}

Charge carrier recombination is one of the most important processes taking place in organic electronic and optoelectronic devices and to a very large extent determines working parameters of the devices. In organic light emitting diodes (OLEDs) the recombination giving photons is a desirable process delivering light, while in organic photovoltaics (OPV) the recombination should be suppressed by all means in order to provide efficient carrier separation and maximal electric power output. Study of recombination is the area of thriving experimental and theoretical endeavor. All recombination processes are divided into two major classes: geminate recombination and bimolecular recombination. For the geminate recombination both carriers are born by the same photon, originate from the same transport site and initially are located close to each other. In this paper we consider only the bimolecular recombination where initial separation and origin of carriers are arbitrary.  Assuming spatially homogeneous distribution of carriers, recombination  kinetics is governed by the equation
\begin{equation}\label{kin}
    \frac{d(n,p)}{dt}=-\gamma np
\end{equation}
where $n(t)$, $p(t)$ are the concentration of electrons and holes, correspondingly, and $\gamma$ is the rate constant (we assume that the intrinsic concentration of carriers is negligible).

In many papers it is assumed that the bimolecular recombination is in fact the Langevin recombination with the rate constant
\begin{equation}\label{Lan}
\gamma_L=\frac{4\pi e}{\varepsilon}\left(\mu_{+}+\mu_{-}\right)
\end{equation}
where $\varepsilon$ is the dielectric constant of the medium and $\mu_{+}$, $\mu_{-}$ are mobilities of holes and electrons, correspondingly. This kind of recombination was considered very long ago by Paul Langevin in his pioneer paper \cite{Langevin:433}.  To a very large extent the use of the Langevin rate constant is explained by the lack of detailed knowledge of the charge recombination in amorphous (and, hence, spatially inhomogeneous) semiconductors. At the same time there is a general agreement that the bimolecular recombination is indeed of the Langevin type in OLEDs, as suggested in many experimental papers \cite{Blom:930,Blom:479,Dicker:45203,Pivrikas:125205,Kuik:4502,Kuik:093301,Wetzelaer:165204}.

Nonetheless, in many cases the experimentally measured rate constant is much smaller than the Langevin constant, the so-called reduction factor $\zeta=\gamma/\gamma_L$ could achieve $0.1$, $0.01$, or even $1\times 10^{-4}$ \cite{Lakhwani:557,Proctor:1941,Kniepert:1301401,Pivrikas:176806,Juska:2858,Deibel:163303,Deibel:075203}.
Usually the discrepancy is attributed to the specific mesoscopic inhomogeneous structure of the materials in the device, especially in the case of photovoltaic devices, where the mesoscopically inhomogeneous structure of the material is specially arranged to achieve a better charge separation \cite{Burke:1500123,Proctor:1941}. Typically, such structure is organized by manufacturing of the working material of the OPV device as a mixture of two different materials, one of which serving as electron donor and another one as electron acceptor, thus creating separate pathways for electrons and holes.

Computer simulations support validity of \eq{Lan} for the case of model disordered materials having mesoscopically homogeneous spatial structure and spatially noncorrelated random energy landscape with the Gaussian density of states (DOS) \cite{Albrecht:455,Groves:155205,Holst:235202}, thus supporting the idea of the necessity of the large scale inhomogeneity for the significant reduction of $\zeta$. In this paper we are going to demonstrate that the real polar amorphous organic  semiconductors which have spatially correlated random energy landscape could demonstrate strongly suppressed recombination and very small $\zeta$ factor without mesoscopic inhomogeneity and the decrease of $\zeta$ is directly related to the long range spatial correlation of the random energy landscape.

\section{Effective charge of the trapped carrier in amorphous polar organic material}

We consider the recombination of carriers in mesoscopically spatially homogeneous polar amorphous semiconductors where the dominant part of the total energetic disorder is provided by randomly located and oriented permanent dipoles. For the high concentration of dipoles the DOS has the Gaussian shape \cite{Dieckmann:8136,Novikov:877e} and slow spatial decay of the electrostatic potential of the individual dipole leads to the long range spatial correlation of the resulting random energy landscape $U(\vec{r})$ being the sum of the electrostatic contributions of all dipoles.\cite{Novikov:14573,Dunlap:542} The model of the exponential DOS is sometimes considered as a worthy alternative to the model of the Gaussian DOS, especially for tails of the DOS associated with deep traps. \cite{Blom:930,Blom:479,Vissenberg:12964,Kuik:093301,Nicolai:195204} Nonetheless, various incarnations of the Gaussian DOS model are, probably, the most popular models for description of the transport properties of amorphous organic semiconductors, successfully explaining many general features of hopping charge transport.\cite{Bassler:15,Novikov:2584,Novikov:2532} For this reason we limit our consideration to the case of the Gaussian DOS naturally arising in the dipolar amorphous organic materials.

Let us consider the case where mobilities of the opposite kinds of carriers are very different (e.g., because of the large difference of the intermolecular transfer integral): let, for example, $\mu_+\ll\mu_-$. In addition, we consider the case where concentration of carriers is low, inter-carrier distance is large and typical time before recombination is long enough. In this case we may assume that carriers have enough time to undergo the full energetic and spatial relaxation before the recombination event. Hence, the mobile carriers move with mobilities determined by the energetic disorder of the medium and slow positive carriers mostly dwell in the deep valleys of the energy landscape $U(\vec{r})$. If so, then the recombination process could be considered as a recombination of mobiles electrons with almost static holes. We assume also that the applied electric field is negligible. Low density of carriers means that we consider either the special case where the initial density of carriers is low enough or the later stage of high density recombination when the majority of carriers already recombined.

\begin{figure}[tbp]
\includegraphics[width=3.25in]{figure1a.eps}
\vskip10pt
a)

\medskip
\includegraphics[width=3in]{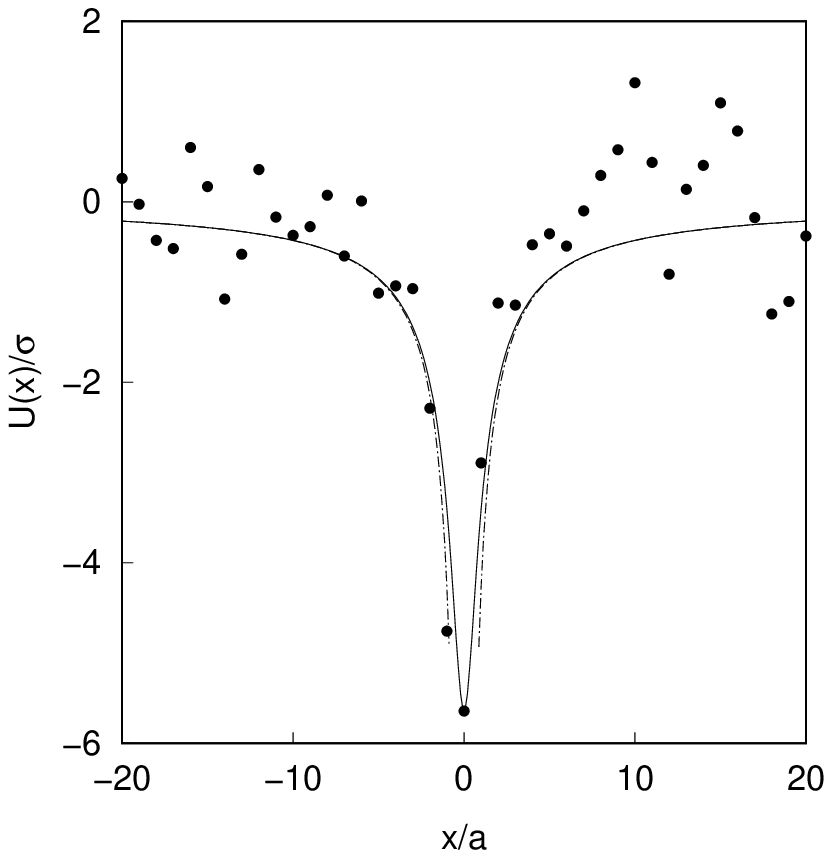}
\vskip10pt
b)
\caption{a) Two dimensional cross-section $z=0$ of the 3D sample of the dipole glass in the vicinity of the deep valley ($U_0=5.6\sigma$) located at $x=y=z=0$, black dots show the energies of neighbor sites, and $a$ is the lattice scale. Surface created by lines shows the averaged energy $-U_0 C(x,y,0)/C(0,0,0)$. b) One dimensional cross-section $z=y=0$ of the same valley, the solid line again shows $-U_0 C(x,0,0)/C(0,0,0)$, the dotted line shows the asymptotics $-U_0 Aa/|x|$.}  \label{valley}
\end{figure}

Let us consider the recombination of electron with the particular hole located at $\vec{r}=0$ in a valley having minimal energy $U(0)=-U_0$. Electron is attracted to the trapped hole by the combined effect of the bare Coulomb attraction to that hole and dipolar contribution from the potential well localizing the hole. Our crucial approximation is the replacement of the exact fluctuating dipole potential energy $U(\vec{r})$ around positive charge by its average value (see \fig{valley}). A similar approach was used by Nikitenko \latin{et al.}\cite{Nikitenko:7776} for the analysis of charge carrier transport in polar amorphous organic materials. We mean here the conditional average taking into account the exact value $U(0)=-U_0$. For the random Gaussian landscape this conditional average is exactly equal to $-U_0 C(\vec{r})/C(0)$, where $C(\vec{r})=\left<U(\vec{r})U(0)\right>$ is the binary correlation function, angular brackets mean the statistical averaging over all possible random environments, and $C(0)=\sigma^2$ is the rms energetic disorder \cite{Novikov:14573}. For $r\gg a$, where $a$ is intermolecular distance in the material, $C(\vec{r})\approx A\sigma^2 a/r$ (practically, this relation is valid with good accuracy even for $r/a\simeq 2-3$, see \fig{valley}b) \cite{Novikov:14573,Dunlap:80}. Typically, $A\simeq 1$; for example, for the lattice model of the disordered polar material where randomly oriented dipoles occupy sites of the simple cubic lattice  (the so-called dipole glass model) $A=0.76..$ \cite{Dunlap:80} (we use this very value of $A$ for all further estimations). Hence, for a large distance the complex hole+dipoles provides the potential
\begin{equation}\label{energy}
    \varphi(\vec{r})=\frac{e}{\varepsilon r}-\frac{U_0}{e}\frac{C(\vec{r})}{C(0)}\simeq\frac{e}{\varepsilon r}-U_0\frac{A a}{er}
\end{equation}
and that potential could be considered as a potential generated by the point charge with the effective charge
\begin{equation}\label{charge}
    e^{\ast}=e-U_0 \frac{A a \varepsilon}{e}
\end{equation}
We may assume that the elementary act of the bimolecular recombination in the amorphous semiconductor could be considered as process somewhat similar to the Langevin recombination in non-disordered medium but taking place between mobile electron with charge $-e$ and static hole having effective charge $e^\ast$. Note that for valleys, where $U_0 > 0$ and where holes dwell most time, $e^\ast < e$ and could even become \textit{negative} for especially deep valleys, thus resulting in the effective \textit{repulsion} between electron and hole. This does not mean that the recombination between electrons and deeply trapped holes becomes impossible, the diffusive motion of the electrons could still eventually bring charges close to each other leading to recombination, but the corresponding rate constant should be severely diminished. Obviously, for such deeply trapped holes the recombination process becomes very different from the usual Langevin recombination.

\section{Recombination rate constant: general expression}

For the calculation of the rate constant for a single hole with the particular value of $U_0$ we are going to use the method of  Smoluchowski and Debye (see \onlinecite{Rice:book}), where the rate constant is defined through the stationary solution for $t\to\infty$ of the equation for the probability density $\rho(r,t)$ for the mobile carrier
\begin{equation}\label{eq_SD}
\frac{\partial \rho}{\partial t}=\frac{D_{-}}{r^2}\frac{\partial}{\partial r}\left[r^2\left(\frac{\partial\rho}{\partial r}+\beta\frac{\partial U}{\partial r}\rho\right)\right]
\end{equation}
where $\beta=1/kT$ and $D_{-}$ is the diffusivity of electrons.  Here and later $U(\vec{r})=-ee^{\ast}/\varepsilon r$ and we take into account the spherical symmetry of the problem. In addition we have the boundary conditions: 1) $\rho(R,t)=0$, meaning the instant recombination of the pair separated by the distance $R$, and 2) $\rho(\infty,t)=1$ which is appropriate for the inexhaustible reservoir of mobile carriers. The stationary density $\rho_s(r)=\rho(r,t\to\infty)$ obeys the equation
\begin{equation}\label{eq_SD_stat}
\frac{d }{d r}\left[r^2\left(\frac{d\rho_s}{dr}+\beta\frac{dU}{dr}\rho_s\right)\right]=0
\end{equation}
and the recombination rate constant for that particular hole is defined by the total flux of mobile carriers through the absorbing sphere of radius $R$
\begin{equation}\label{k_SD}
\gamma(U_0)=4\pi D_{-} R^2\left.\frac{\partial \rho_s}{\partial r}\right|_{r=R}
\end{equation}
Solution of \eq{eq_SD_stat} is
\begin{eqnarray}\label{Wst}
\rho_s(r)=\exp\left(-\beta U\right)\left[1-S(r)/S(R)\right]\\
S(r)=\int\limits_r^\infty \frac{dz}{z^2}\exp\left(\beta U\right)
\end{eqnarray}
In our case $U(\vec{r})=-ee^{\ast}/\varepsilon r$, so
\begin{equation}\label{S(r)}
S(r)=\frac{kT\varepsilon}{ee^{\ast}}\left[1-\exp\left(-\frac{ee^{\ast}}{kT\varepsilon r}\right)\right]
\end{equation}
and
\begin{equation}\label{k_1}
\gamma(U_0)=\frac{4\pi D_{-}}{S(R)}
\end{equation}
This is the rate constant for the particular case where the positive charge is located at the bottom of the valley with depth $U_0$. If we assume the quasi-equilibrium distribution of static charges after the full relaxation, we obtain the full rate constant by the averaging of the rate constant $\gamma(U_0)$ with the density of occupied states
\begin{equation}\label{P_occ}
P_{\rm occ}(U_0)=\frac{1}{(2\pi\sigma^2)^{1/2}}
\exp\left[-\frac{(U_0-U_\sigma)^2}{2\sigma^2}\right]
\end{equation}
where $U_\sigma=\sigma^2/kT$ and after simple transformations we obtain for the full recombination rate constant
\begin{equation}\label{k_SD1}
\gamma=\left<\gamma(U_0)\right>=\frac{4\pi D_{-} R}{\left(2\pi\delta^2\right)^{1/2}}
\int\limits_{-\infty}^\infty dy \frac{y}{\exp(y)-1}
\exp\left[-\frac{(y-y_s)^2}{2\delta^2}\right]
\end{equation}
where $y_\sigma=\left(\frac{\sigma}{kT}\right)^2\frac{Aa}{R}$, $y_c=R_{\rm Ons}/R$, $y_s=y_\sigma-y_c$, $\delta=\frac{\sigma}{kT}\frac{Aa}{R}$, $R_{\rm Ons}=e^2/\varepsilon kT$ is the Onsager radius and we should expect $R\simeq a$. Cut-off for $y > 0$ provided by the exponent in the denominator of the fraction in the integral in \eq{k_SD1} describes the drastic decrease in the recombination rate when  the effective charge $e^{\ast}$ of the static hole becomes negative, thus providing the repulsion between the hole and approaching electrons.

\Eq{k_SD1} is valid if we assume that the quasi-equilibrium distribution of holes described by \eq{P_occ} is permanently maintained irrespective of recombination for all times (thus, we assume that holes are not totally immobilized in deep valleys of the random energy landscape).  Obviously, it could be possible only for slow recombination when drop of electron concentration $\Delta n=n(t)-n(t+\tau_{\rm rel})$ for the time interval equal to the hole relaxation time $\tau_{\rm rel}$ is small compared to $n(t)$ (assuming $n=p$)
\begin{equation}\label{quasi}
\frac{\Delta n}{n}\simeq\frac{\gamma n^2 \tau_{\rm rel}}{n} =\gamma n \tau_{\rm rel}\ll 1
\end{equation}
We see that for low concentration of carriers this inequality could be fulfilled for any $\tau_{\rm rel}$, though the actual concentration of carriers where our consideration becomes accurate strongly depends on $\tau_{\rm rel}$. For low temperature or strong disorder $\tau_{\rm rel}$ could be large,\cite{Bassler:15} but these very conditions lead to small $\gamma$, thus making inequality in \eq{quasi} less restrictive.

We may effectively take into account a slow quasi-geminate recombination of carriers at a short distance described by the rate constant $k_g$ by applying a different boundary condition at $r=R$, i.e. equating the rate of the slow quasi-geminate recombination to the total flux through the sphere of the radius $R$
\begin{equation}\label{k_gem}
k_g\rho(R,t)=4\pi D_{-} R^2\left.\left[\frac{\partial \rho}{\partial r}+\frac{\rho}{kT}\frac{\partial U}{\partial r}\right]\right|_{r=R}
\end{equation}
(the so-called radiation boundary condition \cite{Rice:book}). In this case the final expression for the resulting recombination rate constant is
\begin{equation}\label{k_SD2}
\gamma=\frac{4\pi D_{-}R}{\left(2\pi\delta^2\right)^{1/2}}
\int\limits_{-\infty}^\infty dy \frac{y}{\left(\lambda y+1\right)\exp(y)-1}
\exp\left[-\frac{(y-y_s)^2}{2\delta^2}\right]
\end{equation}
with $\lambda=4\pi D_{-}R/k_g$ and goes to \eq{k_SD1} for the case of instant quasi-geminate recombination $k_g\to\infty$.

Probably, a good estimation for the rate constant for the general case of an arbitrary relation between $D_{-}$ and $D_{+}$ could be provided by the replacement $D_{-}\Rightarrow D_{+}+D_{-}=D$ in eqs \ref{k_SD1} and \ref{k_SD2}. This replacement correctly captures both limit cases $D_{+}/D_{-}\to 0$ and $D_{-}/D_{+}\to 0$ and provides a reasonable interpolation for the arbitrary ratio of $D_{-}$ and $D_{+}$. In addition, for the case of negligible disorder $\sigma/kT\to 0$ the Gaussian in \eq{k_SD1} goes to the delta-function $\delta(y-y_c)$ and the rate constant becomes
\begin{equation}\label{k_SD1a}
\gamma=\frac{4\pi D_{-}y_c R}{1-\exp(-y_c)}
\end{equation}
which in the limit $y_c \gg 1$ with the suggested replacement $D_{-}\Rightarrow D$ and assuming the validity of the Einstein relation $\mu_{\pm}=eD_{\pm}/kT$ gives exactly the usual Langevin rate constant $\gamma_L$.

Our approach for incorporation of the slow quasi-geminate recombination is similar to that developed by Hilczer and Tachiya \cite{Hilczer:6808}, though they considered the recombination in non-disordered medium. Naturally, in the limit of vanishing disorder $\sigma\to 0$ we reproduce theirs result for the rate constant.

\section{Comparison with experiment, computer simulation, and other theoretical models}
\label{SectCompar}

According to \eq{k_SD2}, the most general formula of our consideration, the recombination rate constant has the form
\begin{equation}\label{k_SD3a}
\gamma=\gamma_R F(y_s,\delta,\lambda)
\end{equation}
where $\gamma_R=4\pi DR$ and $F$ is a dimensionless function of 3 dimensionless parameters $y_s$, $\delta$, and $\lambda$. Rich structure of \eq{k_SD2} suggests the possibility of a wide variety of recombination regimes depending on particular values of $y_s$, $\delta$, and $\lambda$. Thorough analysis of \eq{k_SD2} and calculation of $\gamma$ for various cases is provided in the Appendix, where we try to cover the whole physically meaningful range of parameters. In this Section we consider the behavior of $\gamma$ for typical values of parameters relevant for amorphous organic semiconductors.

\begin{figure}[tbp]
\includegraphics[width=3in]{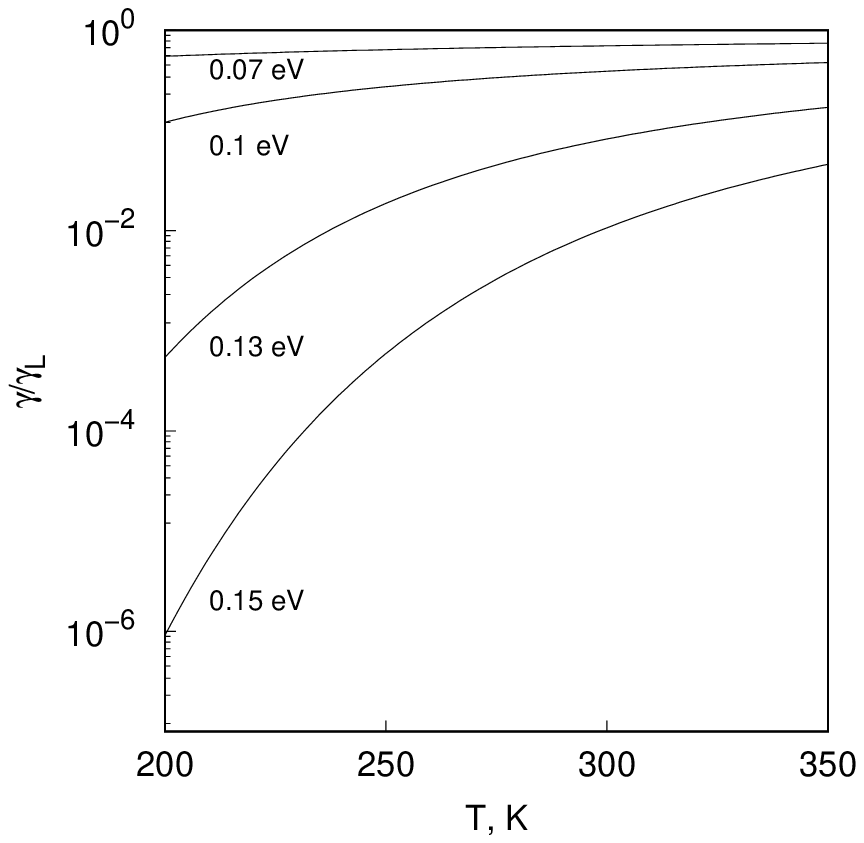}

a)

\medskip
\includegraphics[width=3in]{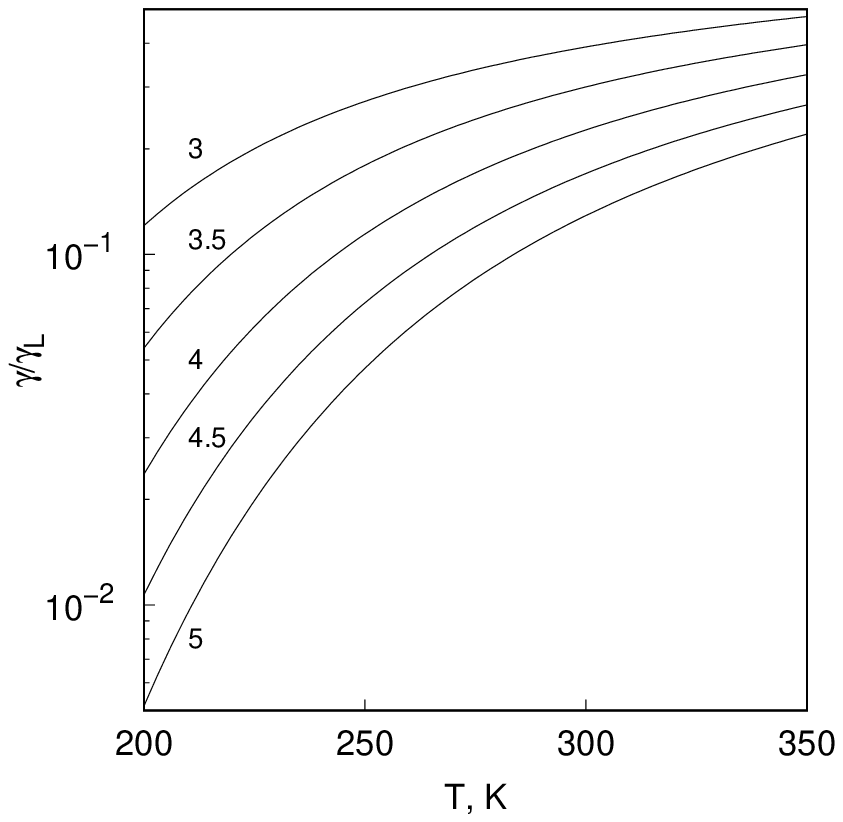}

b)
\caption{Deviation of the bimolecular recombination rate constant from the Langevin constant according to \eq{k_SD1} (hence, we assume the instant quasi-geminate recombination with $\lambda=0$). a) Solid lines show the ratio $\gamma/\gamma_L$ for various values of $\sigma$, indicated near the corresponding curve. For other relevant parameters the typical values $a=R=1$ nm and $\varepsilon=3$ have been used. b) Plot of the ratio $\gamma/\gamma_L$ for various values of $\varepsilon$, indicated near the corresponding curve, and the same values of $a$ and $R$. Increase of $\varepsilon$ is analogous to the increase of $\sigma$ because it again strengthens effect of the disorder (assuming the constant $\sigma=0.1$ eV).}
\label{real}
\end{figure}

Let us start with the estimation of $y_s$ and $\delta$. Numerous experimental transport studies suggest that $\sigma$ falls in the range $0.05-0.15$ eV, and the typical experimental temperature varies from 200K to 350K (most experiments have been carried out around room temperature) \cite{Bassler:15,Borsenberger:9,Novikov:2584}. In amorphous organic semiconductors typically $a\simeq R\simeq 1-1.5$ nm, $\varepsilon=3-4$, so $y_c\simeq 10-30$, $\delta\simeq 2-7$, $y_s\simeq -20-30$. This estimation means that real organic semiconductors could possibly demonstrate various recombination regimes, but typically those regimes fall in the class "broad Gaussian regime" with $\delta\gg 1$. Unfortunately, in the majority of amorphous organic semiconductors  realization of the condition $(y_s-\delta^2)/\delta \gg 1$ is almost impossible. This condition could be fulfilled either at very low temperature (where experimental measurements are very difficult or even impossible) or for materials with high $\varepsilon\simeq 4.5-5$ \cite{Vannikov:K47} and large $a\simeq 2-3$ nm.\cite{Young:388} Such extreme values of $\varepsilon$ and $a$ are very unusual. For this reason the most appropriate way to calculate $\gamma$ for real semiconductors and typical experimental conditions is the use of \eq{erfc} or direct numerical evaluation of the integral in \eq{k_SD2}. Typical behavior of $\gamma/\gamma_L$ for reasonable values of $\sigma$ and $\varepsilon$ is shown in Figure \ref{real}.

A possible way to verify our results could be a comparison with the results of computer simulation. Unfortunately, there are no papers considering the simulation of charge carrier recombination in amorphous dipole medium, but there are papers simulating the recombination in the Gaussian uncorrelated random energy landscape (B{\"a}ssler's Gaussian Disorder Model (GDM) \cite{Bassler:15}). Comparison of our results with the simulation data for the GDM gives an excellent possibility to verify a very essence of our approach. Indeed, for the GDM the correlation function $C(\vec{r})\propto \delta(\vec{r})$ and \eq{energy} immediately tells us that the long range behavior of the potential of the trapped charge is not modified, $e^\ast=e$, so $\gamma\approx \gamma_L$. This very behavior was indeed observed in simulations \cite{Albrecht:455,Groves:155205,Holst:235202}. Hence, small $\zeta$ in mesoscopically homogeneous amorphous semiconductors is a direct manifestation of the correlated nature of the energy landscape.

There is a seeming disagreement between our results and the experimental data for OLEDs, where authors concluded that the bimolecular recombination is the Langevin one  \cite{Blom:930,Blom:479,Dicker:45203,Pivrikas:125205,Kuik:4502,Kuik:093301,Wetzelaer:165204}. We see several reasons why the significant deviation from the Langevin recombination does not occur (i.e., why $\zeta_{\rm exp}\simeq 1$). We believe that the most important reason is a very indirect way to extract the bimolecular recombination rate constant $\gamma_{\rm exp}$ from the experimental data. Direct observation of the decay kinetics $p(t)$ or $n(t)$  is impossible, and extraction of $\gamma_{\rm exp}$ usually invoke many assumptions about details of the transport mechanism, trap distribution, etc. Typical examples are provided in refs \citenum{Blom:930} and \citenum{Pivrikas:125205} where meticulous description of the procedure to estimate $\gamma_{\rm exp}$ is described. Hence, we have to consider all values of $\gamma_{\rm exp}$ obtained in refs \citenum{Blom:930,Blom:479,Dicker:45203,Pivrikas:125205,Kuik:4502,Kuik:093301,Wetzelaer:165204} with great care, they could easily deviate from the true rate constants by one or two orders of magnitude. Indeed, detailed description in refs \citenum{Blom:930} and \citenum{Pivrikas:125205} demonstrates that the best possible accuracy for  $\gamma_{\rm exp}$ and $\zeta_{\rm exp}$ is no better than one order of magnitude.

Additional complication is provided by the contribution of trap-assisted recombination (TAR) which is common in some organic semiconductors \cite{Kuik:093301,Wetzelaer:165204,Lakhwani:557}. Contributions from the trap-assisted and Langevin recombination are not easy to separate, though spectra of the luminescence associated with particular types of recombination are typically  different thus giving the possibility to isolate the individual contributions \cite{Wetzelaer:165204,Kuik:256805}. Our recombination scenario to some extent is close to the TAR, with slow carriers being trapped for a long time in deep valleys of the random energy landscape. The crucial difference is that for the true TAR energy levels of traps are separated by some gap from the energy manifold where charge transport occurs, while in our case there is no such gap. Consideration of the true TAR for the spatially correlated energy landscape could be an interesting and important development of the current study.

Taking into account all these complications it should be very useful to measure $\gamma_{\rm exp}$ in materials having very significant disorder with $\sigma\simeq 0.15$ eV where we should expect very low $\zeta$. At the moment there are no papers reporting reliable measurements of $\gamma$ in strongly disordered organic semiconductors. Nonetheless, in the recent paper\cite{Lee:1800001} it was found that addition of dopants having large dipole moments and, thus, giving noticeable contribution to the total dipolar disorder, leads to the suppression of the Langevin recombination. Obviously, further studies are  needed for the reliable elucidation of the true mechanism of the influence of dipole dopants on charge carrier recombination.

There is another possible reason for the closeness of $\gamma_{\rm exp}$ to $\gamma_L$. We consider here the case of dipolar materials where the long range behavior of the correlation function is of the Coulomb type $C(\vec{r})\propto 1/r$. In fact, many materials considered in refs \citenum{Blom:930,Blom:479,Dicker:45203,Pivrikas:125205,Kuik:4502,Kuik:093301,Wetzelaer:165204} have rather low dipole moments and the dominant part of the random energy landscape is probably generated by randomly located and oriented quadrupoles \cite{Novikov:2584,Novikov:954,Novikov:2532}. For such materials $C(\vec{r})\propto 1/r^3$ and the effect of disorder on the recombination rate constant could not be described by the effective charge $e^\ast$, though the general consideration using Smoluchowski-Debye approach is still  possible. Faster decay of  $C(\vec{r})$ leads to the smaller deviation of $\gamma_{\rm exp}$ from $\gamma_L$, analogously to the case of the GDM. Bimolecular recombination of charge carriers in quadrupolar materials will be considered in a separate paper.

At last, there is yet another reason why $\zeta_{\rm exp}$ is greater than expected; it is associated with the effect of the applied electric field $E$. We consider the recombination for $E=0$ only, while in experiments the recombination constant is estimated for nonzero electric field. Computer simulation indicates that $\zeta$ grows with $E$,\cite{Albrecht:455} thus making the difference between our results and experimental data less drastic.

We have to clarify the difference between our approach and papers of Andriassen and Arkhipov,\cite{Adriaenssens:541,Arkhipov:6869} who also considered the deviation of the bimolecular rate constant from $\gamma_L$ in disordered materials. Our results show that the reason for this deviation is not the disorder \latin{per se}, but the spatial correlation of the disorder. Without correlation $\zeta\approx 1$ irrespectively of the magnitude of the disorder.

\begin{figure}[tbp]
\includegraphics[width=3in]{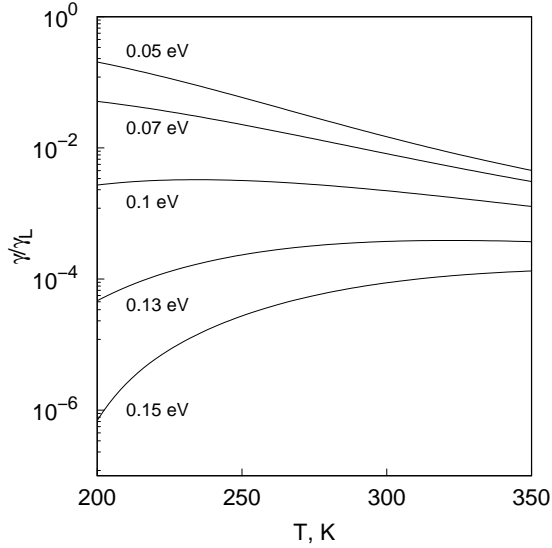}
\caption{Effect of the slow quasi-geminate recombination with $\lambda\gg 1$ on the temperature dependence of the ratio $\gamma/\gamma_L$ for various values of $\sigma$, indicated near the corresponding curve. For other relevant parameters we assume $R=1$ nm, $\varepsilon=3$, $\lambda_0=4\pi D_0 R/k_0=1\times 10^5$, and $E_a=0.2$ eV.}  \label{tachiya}
\end{figure}

As we already noted, our approach could be naturally considered as a far extension of the Hilczer and Tachiya's theory \cite{Hilczer:6808} to the case of disordered medium. We have to admit that the independent estimation of $\lambda$ from first principles is very difficult, in the striking contrast to $y_s$ and $\delta$, thus greatly complicating estimation of the behavior of $\gamma$. Nonetheless, introduction of slow quasi-geminate recombination with $k_g > 0$ gives an inviting possibility to explain the decrease of $\zeta$ with $T$, observed in some experiments.\cite{ Deibel:163303,Juska:1167} Indeed, Figure \ref{real} demonstrates that for $\lambda=0$ $\frac{d\zeta}{dT} > 0$.  At the same time, the sign of $\frac{d\lambda}{dT}$ is arbitrary and depends on the relation between activation energies of $D$ and $k_g$. \Eq{k_SD2bi} hints that if $\lambda$ grows with $T$ fast enough, then $\zeta$ could decrease with $T$. Numerical calculation using \eq{k_SD2} shows that this is indeed so (see \fig{tachiya}). For the calculation we used the simplest activation dependence for $k_g$
\begin{equation}\label{kg_act}
k_g=k_0\exp(-E_a/kT)
\end{equation}
and the proper relation for $D(T)$
\begin{equation}\label{D_act}
D=D_0\exp\left[-\frac{1}{3}\left(\frac{\sigma}{kT}\right)^2\right]
\end{equation}
which is approximately valid in low field limit, as suggested by the renormalization group analysis\cite{Deem:911} and computer simulation \cite{Novikov:275}.

In our case the small or moderate disorder is clearly favorable for the emergence of the recombination regime with $\frac{d\zeta}{dT} < 0$. We can see that even more complicated behavior is possible, namely the change of the sign of the derivative $\frac{d\zeta}{dT}$. In fact, the very motivation of Hilczer and Tachiya to develop their approach was to explain the decrease of $\zeta$ with $T$ in some organic semiconductors. Unfortunately, invocation of the slow quasi-geminate recombination for the explanation of this phenomenon is not fully justified, mostly because semiconductors in question are typical mesoscopically inhomogeneous materials specifically developed for the OPV applications.\cite{ Deibel:163303,Juska:1167} General properties of the charge carrier recombination in such materials are still barely known. For this reason \fig{tachiya} is not provided for the explanation of the behavior of $\zeta(T)$ in any particular organic semiconductor but just to illustrate the possibility to obtain $\frac{d\zeta}{dT} < 0$ for reasonable values of relevant parameters even in mesoscopically spatially homogeneous amorphous semiconductors.

\section{Conclusions}

We calculate the rate constant  of the bimolecular charge carrier recombination in polar amorphous organic semiconductors in the limit of low applied electric field. We show that the long range spatial correlation of the random energy landscape typical for such materials leads to the deviation of the bimolecular recombination from the Langevin-like process with the resulting rate constant $\gamma$ being in some cases much smaller than the corresponding Langevin rate constant $\gamma_L$.

The most important qualitative conclusion is that the stronger is the  correlation, the lower is the ratio $\gamma/\gamma_L$ (factor $\zeta$), and for the total absence of any spatial correlation (the GDM case) our approach gives $\zeta\approx 1$ in agreement with computer simulations \cite{Albrecht:455,Groves:155205,Holst:235202}. We may expect that in nonpolar amorphous organic materials the deviation of $\gamma$ from $\gamma_L$ is much less pronounced due to faster decay of the disorder correlation function.

We show that small $\zeta$ factor could be achieved in mesoscopically homogeneous amorphous semiconductors with large $\sigma$ and $\varepsilon$ and could not be unambiguously related to the inhomogeneous structure of the organic material.

Suggested approach could be extended to consider the bimolecular recombination in nonpolar amorphous organic semiconductors where the dominant part of the random energy landscape is provided by quadrupole molecules or to the case of trap-assisted recombination in semiconductors  with spatially correlated energy landscape.

\begin{acknowledgement}
Financial support from the FASO State Contract No. 0081-2014-0015 (A.N. Frumkin Institute) and Program of Basic Research of the National Research University Higher School of Economics is gratefully acknowledged.
\end{acknowledgement}


\section{Appendix. Recombination rate constant: various limit cases}

\renewcommand{\theequation}{A\arabic{equation}}
\setcounter{equation}{0}

\Eq{k_SD2} demonstrates rich and complex structure hinting for the possibility of many different recombination regimes. Actual dependence of the recombination rate constant $\gamma$ on relevant physical parameters $T$, $\sigma$, $\varepsilon$ and others for any particular case  is determined by the relation between values of the dimensionless parameters $y_s$, $\delta$, and $\lambda$. Typical values of these parameters (and, hence, typical dependences of $\gamma$) for amorphous organic semiconductors are discussed in Section \ref{SectCompar}. In this Appendix we consider the much broader range of possibilities, some of them cannot be realized in today's semiconducting materials. Nonetheless, we believe that this consideration is not worthless and some recombination regimes, though not feasible today, might  be observed in semiconducting materials developed in future.

Hence, we consider here as many physically meaningful regimes as possible, not limiting our attention to the particular values of $y_s$, $\delta$, and $\lambda$, typical for organic semiconductors: we deal here with the general case of the amorphous material having the spatially correlated Gaussian DOS with the dipolar-like correlation function, and the only necessary conditions are $\delta \ge 0$ and $\lambda \ge 0$.

Analytic calculation of the rate constant $\gamma$ in \eq{k_SD2} in the general case is not possible. Let us consider various limit cases which could be treated analytically.

\subsection{The case of sharp Gaussian $\delta\ll 1$}

The simplest tractable limit is $\delta \ll 1$, where the Gaussian in  \eq{k_SD2} goes to the delta function and
\begin{equation}\label{k_SD_d0}
\gamma=\frac{\gamma_R y_s}{\left(\lambda y_s +1\right)\exp(y_s)-1}
\end{equation}
The most reasonable case of small $\delta$ is provided by the negligible disorder $\sigma/kT\to 0$, and in this case the resulting rate constant for $R_{\rm Ons}/R\gg 1$ goes to the usual Langevin constant $\gamma_L$ (for $\lambda =0$).

More exotic possibility is the case of strong disorder $\sigma/kT\gg 1$ and huge recombination radius $a/R\ll 1$, so that $\delta \ll 1$ but still $y_s\gg 1$. In this case the rate constant in \eq{k_SD_d0} still has an exponentially strong dependence on the effective disorder $\sigma/kT$. We have to admit that at the moment we cannot present any concrete organic semiconductor demonstrating such behavior.

\subsection{The case of broad Gaussian $\delta\gg 1$}

Let us consider the opposite case $\delta\gg 1$. When this inequality is valid, then the most natural situation is that for the position of the maximum of the Gaussian we have $y_s\gg 1$, too. Moreover, typically $y_s\simeq \delta^2$, and ratio of the position of the maximum and width of the Gaussian obeys the inequality $y_s/\delta \gg 1$.  Hence, we may assume that the relevant region of the integration is located far away from $y\simeq 0$ and we may simplify the integral in \eq{k_SD2}
\begin{equation}\label{k_SD2b}
\gamma\simeq\frac{\gamma_R}{\left(2\pi\delta^2\right)^{1/2}}
\int\limits_{-\infty}^\infty dy \frac{y}{\lambda y+1}
\exp\left[-y-\frac{(y-y_s)^2}{2\delta^2}\right]
\end{equation}
Maximum of the Gaussian in this integral is located at $y_s-\delta^2$, so the more exact condition for the validity of the approximate \eq{k_SD2b} is $(y_s-\delta^2)/\delta \gg 1$. We have
\begin{equation}\label{estim}
y_s-\delta^2=\left(\frac{\sigma}{kT}\right)^2\frac{Aa}{R}\left(1-\frac{Aa}{R}\right)-y_c
\end{equation}
Certainly, $R \ge a$ and the most natural choice is $R\approx a$, while $A< 1$, so the combination $y_s-\delta^2$ indeed could be positive, especially at low temperature, even taking into account the negative contribution from the charge-charge interaction.

\begin{figure}[tbp]
\includegraphics[width=3in]{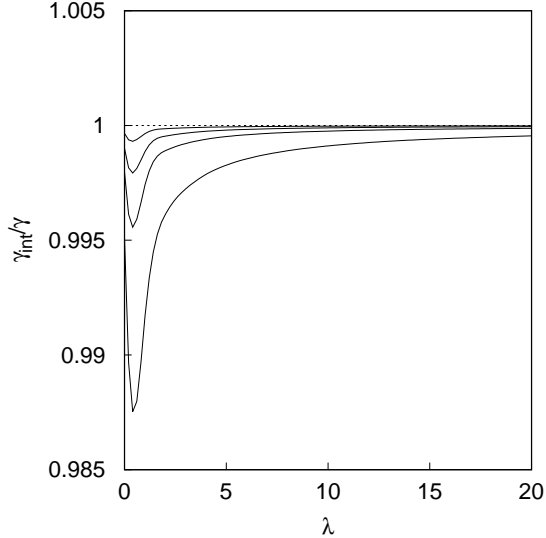}
\caption{Quality of the interpolation \eq{k_SD2bi} is shown. Solid lines show the ratio of $\gamma_{\rm int}$, calculated by \eq{k_SD2bi}, to the rate constant $\gamma$ numerically calculated by \eq{k_SD2} for $y_s$ equals to 20, 25, 30, and 40 from the lowest curve upward, correspondingly. In all cases $\delta=3$.}  \label{inter}
\end{figure}

If $\lambda(y_s-\delta^2) \ll 1$, then the first term in the denominator in \eq{k_SD2b} is not relevant and
\begin{equation}\label{k_SD2b1}
\gamma\simeq \gamma_R\left(y_s-\delta^2\right)
\exp\left(-y_s+\frac{1}{2}\delta^2\right)
\end{equation}
while in the opposite case $\lambda(y_s-\delta^2) \gg 1$
\begin{equation}\label{k_SD2b2}
\gamma\simeq \frac{\gamma_R}{\lambda} \exp\left(-y_s+\frac{1}{2}\delta^2\right)
\end{equation}
A reasonable interpolation between two limits is
\begin{equation}\label{k_SD2bi}
\gamma\simeq \frac{\gamma_R \left(y_s-\delta^2\right)}{1+\lambda\left(y_s-\delta^2\right)}
\exp\left(-y_s+\frac{1}{2}\delta^2\right)
\end{equation}
and the quality of the interpolation formula could be seen in Figure \ref{inter}. Even for not so large $\delta$ and $(y_s-\delta^2)/\delta$ \eq{k_SD2bi} works remarkably well.

\begin{figure}[tbp]
\includegraphics[width=3in]{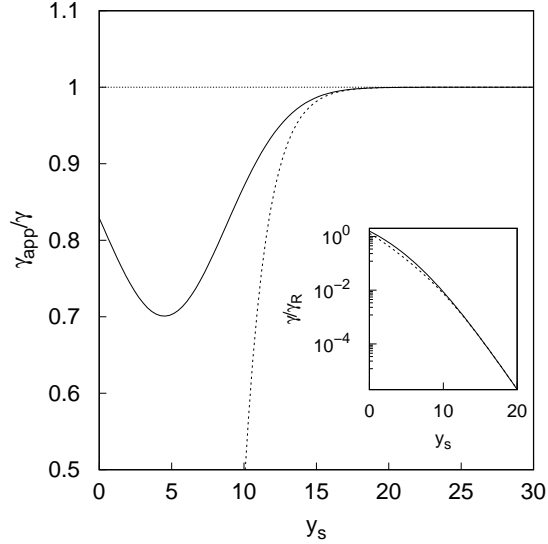}
\caption{Quality of the approximate \eq{erfc} is shown. Solid line shows the ratio of $\gamma_{\rm app}$, calculated by \eq{erfc}, to the rate constant $\gamma$ numerically calculated by \eq{k_SD1} for $\delta=3$. Broken line shows the corresponding ratio for the rate constant $\gamma_{\rm app}$ calculated by \eq{k_SD2b1}. Inset shows the general behavior of the exact (solid line, \eq{k_SD1}) and approximate (broken line, \eq{erfc}) rate constant.}  \label{approx}
\end{figure}

For $\lambda=0$ we may suggest a better approximation than \eq{k_SD2b1} replacing the function under the integral in \eq{k_SD1} by its proper asymptotics
\begin{equation}\label{replace}
\frac{y}{\exp(y)-1}\Rightarrow
\begin{cases}
-y,\hskip5pt y <0\\
y\exp(-y),\hskip5pt y >0
\end{cases}
\end{equation}
the resulting expression takes the form
\begin{eqnarray}\label{erfc}
\gamma\simeq \frac{\gamma_R}{2}\left[\sqrt{\frac{8}{\pi}}\delta\exp\left(-\frac{y_s^2}{2\delta^2}\right)
-y_s
{\rm erfc}\left(\frac{y_s}{\delta\sqrt{2}}\right)+\right.\\
\left.+(y_s-\delta^2)\exp\left(-y_s+\frac{1}{2}\delta^2\right)
{\rm erfc}\left(\frac{-y_s+\delta^2}{\delta\sqrt{2}}\right)\right]\nonumber
\end{eqnarray}
where ${\rm erfc}(x)$ is a complimentary error function. For sufficiently large $y_s$ \eq{erfc} goes to \eq{k_SD2b1}. Quality of approximation could be seen in Figure \ref{approx}. \Eq{erfc} gives a meaningful result even for $y_s=0$, while \eq{k_SD2b1} for $\delta=3$ gives a \textit{negative} rate constant for $y_s < 9$. Yet \eq{erfc} is too cumbersome for the practical use.

More exotic case for $\delta\gg 1$ (strong disorder) is the situation where interaction between carriers is so strong that no matter how large is $y_\sigma$, $y_s$ is still negative and $|y_s| \gg 1$. In this case we may omit the term proportional to $\exp(y)$ in \eq{k_SD2}, so
\begin{equation}\label{k_SD3}
\gamma\simeq -\frac{\gamma_R}{\left(2\pi\delta^2\right)^{1/2}}
\int\limits_{-\infty}^\infty dy y
\exp\left[-\frac{(y-y_s)^2}{2\delta^2}\right]
=\gamma_L-4\pi D a A \left(\frac{\sigma}{kT}\right)^2
\end{equation}
and the rate constant is essentially equal to the Langevin constant with small correction. This is not surprising due to the dominance of the charge-charge interaction over disorder. In this approximation we assume also that the constant $\lambda$ is not unusually large (i.e., quasi-geminate recombination unusually slow), namely $\lambda|y_s|\exp(y_s)\ll 1$. If the opposite is true, then the rate constant obeys \eq{k_SD2b1} but now $y_s$ is negative.

\bibstyle{achemso}
\bibliography{recombination-jpcc-R1}

\end{document}